\documentstyle[epsfig,subfigure,aps]{revtex}
\newcommand{\Dslash}{\rlap{\,/}D}
\newcommand{\dg}{^\dagger}

\newcommand{\lk}{\lambda_k}

\newcommand{\tr}{{\rm tr}}

\newcommand{\lp}{\left(}
\newcommand{\rp}{\right)}
\newcommand{\ls}{\left[}
\newcommand{\rs}{\right]}
\newcommand{\lb}{\left\{}
\newcommand{\rb}{\right\}}
\newcommand{\la}{\left|}
\newcommand{\ra}{\right|}
\newcommand{\bacc}{\begin{array}{cc}}
\newcommand{\ea}{\end{array}}
\newcommand{\ber}{\begin{eqnarray}}
\newcommand{\eer}{\end{eqnarray}}
\newcommand{\AmS}{{\protect\the\textfont2
  A\kern-.1667em\lower.5ex\hbox{M}\kern-.125emS}}

\begin{document}

\title{
Random Matrices and the Convergence of Partition Function Zeros \\ in 
Finite Density QCD
}

\author{M. \'{A}. Hal\'{a}sz$^a$, 
	J. C. Osborn$^b$, 
	M. A. Stephanov$^{c,d}$, 
        J. J. M. Verbaarschot$^b$ }
\address{$^a$Department of Physics and Astronomy,
         University of Pennsylvania, 
                  209 S 33rd Street, Philadelphia, PA 19104, USA \\
         $^b$Department of Physics and Astronomy, 
                  State University of New York, 
                  Stony Brook, NY 11794-3800, USA \\
         $^c$Institute for Theoretical Physics, SUNY,
	          Stony Brook, NY 11794-3840, USA \\
         $^d$Department of Physics, 
                  University of Illinois, 845 W Taylor Street,
                  Chicago, IL 60607-7059, USA}
\draft

\maketitle

\begin{abstract}

We apply the Glasgow method for lattice QCD at finite chemical potential to 
a schematic random matrix model (RMM). In this method the zeros of the
partition function are obtained by averaging the coefficients of
its expansion in powers of the chemical potential.
In this paper we investigate the phase structure by means of Glasgow averaging 
and demonstrate that the method converges to the correct
analytically known result.
We conclude that the statistics
needed for complete convergence grows exponentially with the size of the 
system, in our case, the dimension of the Dirac matrix.
The use of an unquenched ensemble at $\mu=0$ does not give
an improvement over a quenched ensemble.

We elucidate the phenomenon of a faster convergence of certain zeros
of the partition function. 
The imprecision affecting the coefficients of the polynomial in the
chemical potential can be interpeted as the appearance of a spurious phase.
This phase dominates in the regions where the exact partition function
is exponentially small, introducing additional phase boundaries,
and hiding part of the true ones.
The zeros along the surviving parts of the true boundaries remain unaffected. 

\end{abstract}

\section{Introduction}

In contrast with the numerous successes of lattice QCD,
simulations at finite chemical potential \cite{Ko83,Ba86,Ko95,Ba97}
  have oscillated between
mildly promising and outright frustrating. 
The source of the trouble lies in the following.  In the Euclidean 
formulation of QCD, the chemical potential spoils the anti-Hermiticity of the 
Dirac operator. As a result, the fermion determinant is no longer
a real number. In general, it has a complex phase. Hence,
the action cannot serve as a statistical weight in a Monte Carlo sampling
of field configurations. 

Quenched simulations, where the fermion determinant is not
included in the statistical weight, may provide a fairly reliable
approximation to selected observables of the true unquenched theory.
However, in the presence of a chemical potential, quenched simulations have
produced consistently unphysical results \cite{Ko95}. 
The reason is that the quenched theory is the $N_f \rightarrow 0$ 
limit of an unphysical theory 
where the fermion determinant is replaced by its absolute value
\cite{St96,Go88}. This is a theory with a second, ``conjugate'' 
set of anti-quark species together 
with the normal quarks. Because of Goldstone bosons consisting of
a quark and a conjugate anti-quark, the critical chemical potential
in quenched QCD is half the pion mass.
This phenomenon was demonstrated in lattice simulations by Gocksch 
\cite{Go88} using
a $U(1)$ toy model and was understood analytically in \cite{St96} by
using a random matrix model inspired by QCD.

One important conclusion of further studies of the same RMM is that the
phase of the fermion determinant leads to very large cancellations in the
ensemble averaging \cite{Ha97}.
A measure of this phenomenon is given by the fact that the partition function
is proportional to $\exp(-\mu^2 N)$, where $N$ is the size of the random
matrix, corresponding to the number of sites in a lattice simulation.
Cancellations of this magnitude would require prohibitive statistics in
order for a brute force simulation\footnote{where the determinant is included
as an observable in a quenched ensemble} to be successful.
We wish to note that in some models the construction of clever algorithms
makes it possible to deal with these cancellations in
Monte Carlo simulations \cite{chandra}.

It would be hard to overrate the potential importance of a successful
lattice approach to QCD at finite chemical potential. At this time, there is
not even an estimate for the value of the critical chemical potential from
lattice simulations. In the absence of the guidance provided by the lattice,
it is difficult to assess the many semi-empirical descriptions of nuclear
matter at high density \cite{Xa99,Xb99,Kl99,Xd99,Al98,Al97,Sc99}.
For instance, an extension of the RMM to include temperature 
via the first Matsubara frequency
\cite{Ha98} gives reasonable predictions about the phase diagram of QCD,
even while ignoring most of its dynamics.
In the present paper we wish to exploit the qualitative similarity between 
our simple RMM and $N_c=3$ QCD at finite chemical potential in an attempt to 
understand certain lattice results on the problem of finite $\mu$. 

\bigskip

We are interested in the analytic dependence
of the QCD partition function on the chemical potential $\mu$. This can be
obtained by computing the coefficients of the expansion of the partition
function in powers of the chemical potential or the fugacity.
The Glasgow method of lattice QCD \cite{Ba97,Ba99,Lo96,Ba92,Ba90} is designed to do this.
The unquenched partition function can be seen as the quenched average of
the fermion determinant, i.e. averaged only with the gauge action.
In general, one may also use an unquenched ensemble at some fixed chemical
potential $\mu=\mu_0$, and include the inverse fermion determinant at that
same value,
\ber
\label{form1}
Z(\mu)~=~ \langle {\rm det}~\Dslash (\mu) \rangle_{\rm{gauge}}
~=~ C \left\langle \frac{{\rm det}~\Dslash (\mu)}{{\rm det}~\Dslash(\mu_0)} 
\right\rangle_{\rm{gauge},\mu=\mu_0 }~~.
\eer
Here $C$ is an irrelevant constant.
One expects the efficiency of the averaging process to
depend on the overlap between the quantity being averaged and the
distribution used to generate the ensemble.
When these two functions have their largest values in vastly different places,
this is known as an overlap problem. 

In the lattice Glasgow method the fermion determinant is expanded in 
powers of the fugacity $\xi=\exp(\mu)$.
The expansion is finite and exact, since the fermion determinant is just
an $N \times N$ matrix ($N$ is the number of lattice points times the number
of degrees of freedom per site). 
It is obtained by writing 
$\det \Dslash \lp \mu \rp = \xi^N \det \lp P + \xi \rp $,
where $P$ is called the propagator matrix.
The expansion coefficients, written in terms of the
eigenvalues of $P$, are then obtained by ensemble averaging.
The zeros of the partition function in the complex 
$\mu$ plane map out the phase structure.
In particular, the ones close to the real axis define the critical value(s)
of $\mu$.
Since the original paper by Yang and Lee \cite{YL} this approach has been
widely used in statistical mechanics.
For some recent applications we refer to \cite{Shrock}.

With the Glasgow method one obtains information about the full $\mu$-dependence
of $Z$, using an ensemble generated at one fixed value of
the chemical potential.
Of course, the question of the overlap between whatever ensemble one is
using and the fermion determinant for the given $\mu$ remains. 
The issue is even more ominous considering that we are 
unable to even define the notion of an ensemble
at nonzero $\mu$, even in this random matrix model, due to the complex action.

In this paper, we analyze the Glasgow method using a random matrix
model at nonzero chemical potential. As we have already discussed
before, this model mimics the problems of the QCD partition function
at nonzero chemical potential. Since the phase structure of
this model is known analytically, it is an ideal testing ground for
evaluating this algorithm and obtaining a better understanding of its 
problems.

In section 2 we introduce the random matrix model and derive some of
its analytical properties. The bulk of this paper is in section 3. It
contains our numerical analysis of the Glasgow method, and an
explanation is given of the poor convergence of certain zeros of the
partition function. Concluding remarks are made in section 4.

\section{Random matrix model}

We consider a random matrix model (RMM) defined by the partition function 
\cite{St96,Ha97}
\ber
Z(m,\mu) ~&=&~ \int {\cal D} C e^{-N \tr C C^\dagger}
\det \lp D (m , \mu) \rp,
\nonumber\\
D(m, \mu)~&=&~ \ls \bacc m & i C + \mu \\ i C^\dagger + \mu & m \ea \rs
~.
\label{form5}
\eer
Here, $C$ is an $N \times N$ complex matrix, and the integration is
over the Haar measure, 
\ber
\int {\cal D} C~=~
\prod\limits_{ij = 1}^{N}\int dC_{ij} dC_{ij}^{*}~~ .
\eer 
This model was first formulated \cite{SVR,V} for $\mu = 0$ in order to
describe the correlations of the smallest eigenvalues of the 
Dirac operator. In this case it has been shown rigorously that the model 
describes 
the zero-momentum sector of the
low-energy effective partition function of QCD \cite{OTV,DOTV}.  

In the present context, the matrix $D(m,\mu)$
mimics the QCD Dirac operator for quark mass $m$ and chemical potential $\mu$.
The integration over matrix elements replaces the integration over gauge field 
configurations. 
The massless part of our random Dirac operator is anti-Hermitian for 
$\mu=0$, but for $\mu \ne 0$ it has no definite Hermiticity properties.
In order to study the properties 
of the partition function in
the chemical potential plane $\mu$, 
we rewrite the fermion determinant as follows,
\ber
\det \lp D(m,\mu) \rp 
~=~\det \lp \bacc i C + \mu & m \\ m & i C\dg + \mu \ea \rp ~=~
\det \lp F(m) + \mu {\bf 1} \rp~~.
\eer
The matrix $F(m)$ is analogous to the propagator matrix from lattice QCD 
\cite{Gi86,Gi86a,Ba97} in the sense that its eigenvalues are the values of $\mu$ for 
which the fermion determinant vanishes. In terms of 
the eigenvalues $\lk$ of $F(m)$ 
the RMM partition function is 
\ber
Z(m,\mu)~=~\int {\cal D} C e^{-N \tr C C^\dagger}  
\prod\limits_k (\lk + \mu) .
\eer

The quantity $\langle n \rangle~=~\partial_\mu \ln Z(m,\mu)/N$ is
the analog of the baryon number density of QCD. It is equal to 
the ensemble average of 
$\langle \frac 1N \sum_k \frac{1}{\lk + \mu} \rangle$.
In QCD at zero temperature, one expects the baryon number density to be
identically zero for small $\mu$, and then to increase starting from
a certain critical value of $\mu$ \cite{Ha98}.
Similarly, our model shows a phase transition with an increase of the baryon
number density. 
However, $\langle n \rangle$ is not zero below the critical value of $\mu$. 
See \cite{Ha98} for an explanation of the relationship between $\langle n \rangle$
and the baryon number density of QCD.

It is not clear how to define a statistical ensemble of gauge 
field configurations (or random matrices for that matter), 
corresponding to the true partition function with finite $\mu$.
In the quenched approximation one discards the fermion determinant, 
so the partition function does not depend any more on $\mu$ or $m$.
However, a `number density' can still be computed by taking the average of 
$\frac 1N\sum_k \frac{1}{\lk + \mu}$.
Similarly, one can define a quenched `chiral condensate' from the ensemble
average of $\frac 1N {\rm Tr}~\lp D(m,\mu)^{-1} \rp$.
The quenched approximation can be interpreted either
as the limit of a process where one takes the number of flavors
to zero, or the quark mass $m$ to infinity. 
The  quantity 
\ber
n(\mu_0,\mu)~=~\int {\cal D} C e^{-N \tr C C^\dagger } 
\prod\limits_k (\lk + \mu_0) 
\frac 1N \sum\limits_k \frac{1}{\lk + \mu}
\eer
is called the partially quenched baryon number density. The ``sea'' $\mu_0$ 
defines the ensemble, and the ``valence'' $\mu$ probes the eigenvalue
distribution.

\subsection{Quenched eigenvalue distribution}

For $m=0$ the propagator matrix is block-diagonal, and its eigenvalues are
$i$ times those of $C$.
The {\it exact} distribution of the eigenvalues of a general complex 
matrix has been calculated a long time ago by Ginibre \cite{Gi65}.
In our normalization, it is given by
\ber
\label{N_point}
\rho(\lambda_1,\cdots,\lambda_N)~=~{\cal C}
\int d^2\lambda_1 \cdots d^2 \lambda_N e^{- N \sum\limits_k \la \lk \ra^2}
\prod\limits_{k>l} \la \lk - \lambda_l \ra^2 
\eer
The corresponding one-point function is
\ber
\label{1-point}
\rho(\lambda)~=~ {\cal C} e^{- N | \lambda |^2} \sum\limits_{k=0}^{N-1} 
\frac{\lp N |\lambda |^2 \rp ^k }{k!}
\eer
In the large $N$ limit the eigenvalues are uniformly distributed in the
complex unit circle.
This follows from the properties of the truncated exponential, to be discussed
in more detail later.
For $| \lambda | < 1$, the truncated exponential is a good approximation
of the complete one, so $\rho$ is a constant.
For $| \lambda | > 1$, the truncated exponential behaves like a power
of $| \lambda|^2$
which is quickly suppressed by $\exp (- N |\lambda|^2)$, 
so the distribution vanishes with a 
sharp tail of width of order $1/\sqrt{N}$.

For nonzero $m$ we can calculate the eigenvalue distribution of the propagator
matrix in the large $N$ limit using the conjugate replica trick \cite{St96}.
We consider the partition function, where we have replaced the fermion
determinant with its absolute value squared,
\ber
\label{form7}
Z(m,m^*,\mu,\mu^*) ~&=&~ \int {\cal D} C e^{-N \tr C C^\dagger}
\left| \det \lp  \bacc m & i C + \mu \\ i C^\dagger + \mu & m \ea \rp \right|
^{2 \tilde{n}_f}
~.
\eer
Here we can use either $m$ or $\mu$ as a dummy variable probing the eigenvalue
distribution of the corresponding operator as a function of the other variable
which is made real and therefore has a physical meaning.
For the spectrum of the propagator matrix we set $m = m^*$ and probe the
eigenvalues with the complex dummy variable $\mu$.
This is the reverse of what was done in \cite{St96} where the spectrum of
the Dirac operator was investigated for given $\mu$.
The conjugate replica trick also allows us to calculate
the eigenvalue distribution in the present case.

Because of the absolute value of the determinant under the integral,
the partition function is expected to be a smooth function of $\tilde n_f$
and the limit $\tilde{n}_f \rightarrow 0$ should be obtainable from
the partition function for positive integral values of $\tilde n_f$.
For any $\tilde n_f$,
the eigenvalue density is positive definite so it cannot be an 
analytic function of a complex variable. 
The resolvent is defined as
$G(z) = \frac1N \langle \sum_k \frac{1}{\lk - z} \rangle$.
The average eigenvalue density is then given by 
$N Re \lp \partial_{z^*} G(z) \rp$.
For the partition function (\ref{form7}) the resolvent in the $\mu$ plane is 
$G(z=-\mu) = \partial_\mu Z(m,\mu,\mu^*)$.

By standard manipulations \cite{St96} the partition function can be rewritten  
as an integral over an 
$2\tilde{n}_f \times 2\tilde{n}_f$ complex matrix.
For $\tilde{n}_f=1$, one finds
\ber
Z(m,m^*,\mu,\mu^*)~=~\int d^2 a ~d^2 b ~d^2 c ~d^2 d 
~~e^{ - N ( aa^* + bb^* + cc^* + dd^* ) }
\det
\ls \begin{array}{cccc} m + a & \mu & 0 &  id \\ \mu  & m + a^* & ic & 0 \\
0 & i d^* & m^* + b^* & \mu ^* \\ ic^* & 0 & \mu^* & m^* + b \end{array} \rs~~.
\eer
The resulting saddle point equations have two kinds of nontrivial solutions, 
depending on whether the off-diagonal quantities $c,c^*,d,d^*$ 
are identically zero or not. 
If they vanish, the partition function factorizes into pieces that depend
only on $m,\mu$ or on $m^*,\mu^*$.
Then, $G(z)$ is always an analytic function of $z$.
Therefore, the region in the complex plane of $z$ ($m$ or $\mu$) 
where the eigenvalues are located must be dominated by the other kind of
solution, which mixes the parameters and their conjugates.
It turns out that this solution is such that $c=c^*=d=d^*$, and the 
quantity $cc^*$ is positive. For $m=m^*$ real, it is given by
\ber
cc^*~=~\frac{x^2}{x^2 - m^2} - \frac{ m^2}{4(x^2 - m^2)} 
+ \frac{y^2 m^2}{2 x^2 (x^2 - m^2)}
-\frac{ (x^2 + y^2) y^2 m^2 }{4 x^2} -( x^2 + y^2)~~.
\label{eq:mucont}
\eer
The boundary of the domain of eigenvalues is given by the curve, where the
two types of solutions merge, i.e. the condition $cc^*=0$.
One can see immediately that for $m=0$ the boundary $cc^*=0$ 
reduces to the unit circle.

\begin{figure}[t]
  \centering
  \subfigure[$m=0$]{
    \psfig{file=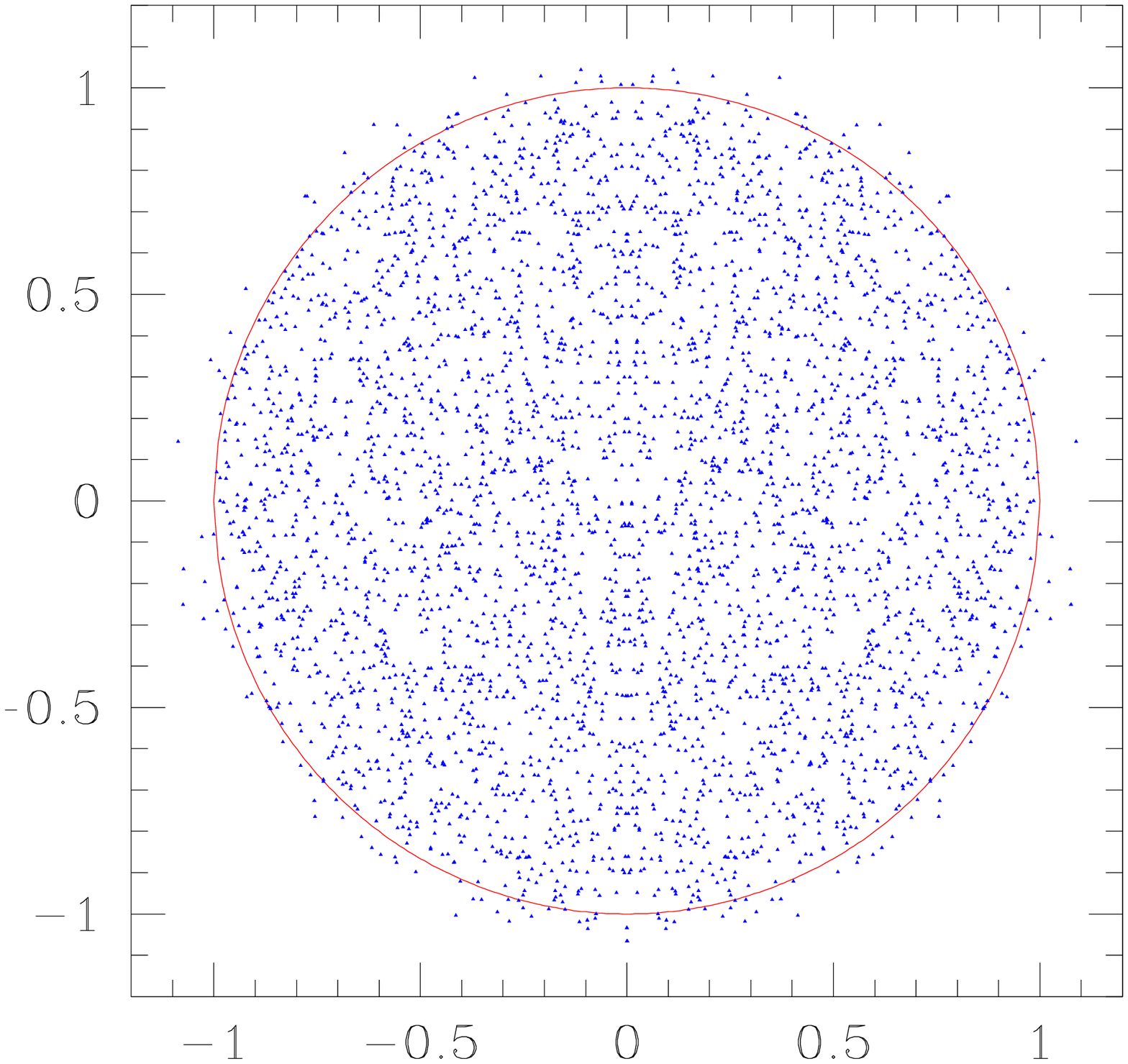,width=8.5cm}
    \label{mucont0}}
  \subfigure[$m=0.0625$]{
    \psfig{file=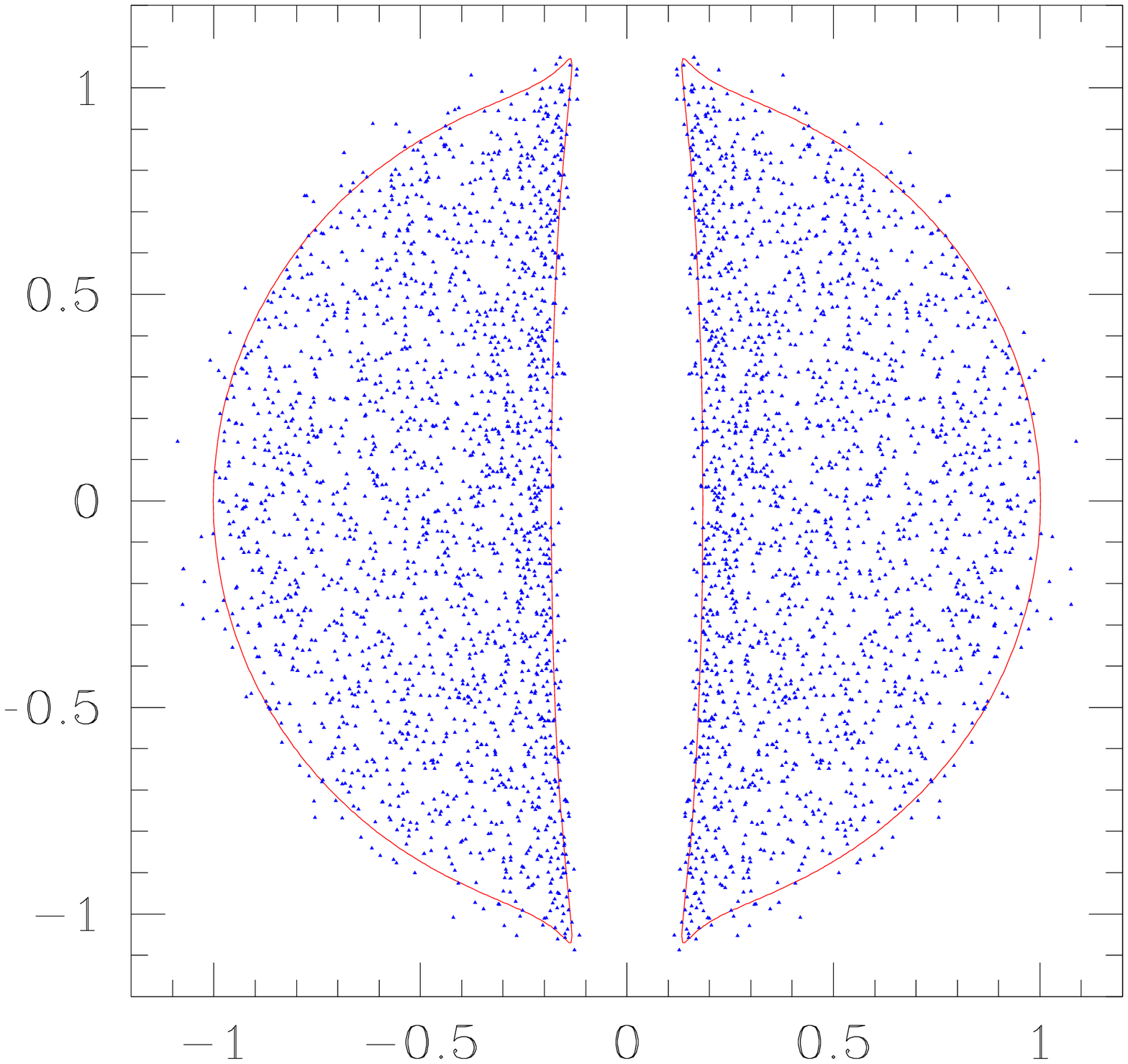,width=8.5cm}
    \label{mucont0625}}
  \caption{Distribution of the eigenvalues in the complex plane of the
	   propagator matrix of size
           $N=96$ along with the analytic curve for the boundary
           given by $cc^*=0$ in (\ref{eq:mucont}).}
  \label{mucont}
\end{figure}

In Fig.~\ref{mucont} we show the distribution of the eigenvalues in the 
complex plane of the
propagator matrix of size $N=96$ for masses of value $m=0$ and $m=0.0625$.
Each plot consists of eigenvalues from 20 independent configurations.
The analytic curve given by $cc^*=0$ is also drawn in each case.
We clearly see that this does give the correct result for the boundary of
eigenvalues.

\subsection{Unquenched partition function}

The unquenched partition function defined in eq.~(\ref{form5}) 
can be computed analytically \cite{JV,St96,Ha97}.
It is given by
\ber
Z(m,\mu) = \int d\sigma d\sigma^\dagger {\rm e}^{- N \tr \sigma \sigma^\dagger}
\det \lp \bacc m + \sigma & \mu \\ \mu & m + \sigma^\dagger \ea \rp^N
\eer
where the integration is over the $N_f \times N_f$ matrix $\sigma$.
In the $N \rightarrow \infty$ limit the integrals can be evaluated via
saddle point approximation. The saddle points are given by a cubic equation,
\ber
\sigma^*~&=&~\sigma \nonumber \\
\sigma (m + \sigma )^2 - \mu^2 \sigma ~&=&~ m + \sigma~~,
\eer
where the matrix variable is diagonal and $\sigma$ is now a number.
For fixed real $m$, there are four branch points 
in the complex $\mu$ plane, given by four of the six $\mu$ values for which
the discriminant of the above equation,
\ber
D~=~ \frac{1}{27} \lp m^4 \mu^2 - m^2 \lp
2 \mu^4 - 5 \mu^2 - \frac{1}{4} \rp + \lp 1 + \mu^2 \rp^3 \rp ~~,
\eer
vanishes. 
The branch points are connected by two branch cuts. The derivatives
of the partition function are discontinuous across these cuts. The points 
where the cuts cross the real axis are interpreted as critical values of 
the parameters.
They indicate a first-order phase transition in the thermodynamic limit 
\cite{St96,Ha97,Ha98}.

For finite $N$, the partition function can be evaluated as a polynomial 
in either $m$ or $\mu$.
The coefficients are obtained by expanding the determinant
and performing the integrals.
One exact expansion of the partition function is given by \cite{Ha97}
\ber
\label{form14}
Z(m,\mu)~=~ \frac{\pi N!}{N^{N+1}}
\sum\limits_{k=0}^{N} \sum\limits_{j=0}^{N-k}
\frac{(N m^2 )^k}{(k!)^2} \frac{(-N \mu^2)^j}{j!} \frac{(N-j)!}{(N-j-k)!}~~.
\eer
The zeros of this polynomial can be readily calculated. 
They are located along lines in the complex plane and
in the limit $N \rightarrow \infty$ they converge 
to the exact cuts found by the saddle point analysis below (\ref{sp}).

\subsection{Phase structure and zeros of a partition function}

When a partition function has a nontrivial 
phase structure in the thermodynamic limit,
the complex plane of one of its thermodynamic parameters
is split into regions separated by cuts.
Inside each region, the partition function is analytic in the parameters, 
so that its derivatives are continuous. 
The (first-order) transitions occur across 
the cuts, where the (logarithmic) derivatives of the partition function will
in general have discontinuities.

Before taking the thermodynamic limit (i.e. at finite $N$), 
the partition function is an analytic function of the parameters.
If this partition function at finite $N$ is a polynomial in any
of its thermodynamic parameters, 
such as in our case the mass or the chemical potential, 
it will have zeros in the complex plane of these parameters. Its logarithmic
derivatives will have poles at the locations of the zeros. 
In the thermodynamic limit, the zeros coalesce into the same 
cuts that define the phase structure.

Our partition function illustrates nicely the connection between the zeros
and the cuts. For $m=0$ it is a truncated exponential,
\ber
Z_N(0,\mu)~=~\frac{\pi N!}{N^{N+1}} \sum\limits_{j=0}^N 
\frac{(-N\mu^2)^j}{j!}~~.
\eer
For $\mu^2  \ll 1$ 
the largest term in this sum occurs well before the truncation 
so we have a good approximation
of the exponential. On the other hand, for $\mu^2 \gg 1$ the series is 
dominated by the term with $j=N$. 
One can obtain a better estimate using the incomplete gamma function, which is
closely related to the truncated exponential \cite{GrRy}:
\ber
n! e^{-x} \sum\limits_{k=0}^n \frac{x^k}{k!}~&=&~
\Gamma(1+n,x) ~=~ \int\limits_{x}^{\infty} e^{-t} t^n dt~~.
\eer
For real positive arguments the situation is simple. If $x$ is less
than the value $t=n$ which maximizes the integrand, then the saddle 
point is integrated over, the integral is a good approximation to the gamma
function and the exponential is obtained. 
For $x > n$ the integral is dominated by its endpoint and the  
integrand is well approximated by the last term of the series.  

If $x$ is complex, the saddle point dominates if the integration contour can
be deformed across the saddle point.
For our partition function
\ber
Z_N(0,\mu)~&=&~ \pi e^{-\mu^2 N} \int\limits_{-\mu^2}^\infty e^{-N u} u^N du~,
\eer
this happens if ${\rm  Re} (\mu^2+\ln\mu^2) < -1$ and ${\rm Re} \mu^2 > -1$.
In this case $Z_N(0,\mu) \sim \exp(-\mu^2 N -N)$.
If the integral is dominated by the endpoint then
$Z_N(0,\mu) \sim \mu^{2N}$. Nontrivial zeros of the
partition function are obtained for values of the chemical potential
where the saddle point contribution and the end point contribution are
of comparable order of magnitude. The zeros are thus given by the equation
\cite{Ha97}
\ber
{\rm Re} (1+\mu^2 +\log \mu^2)=0.
\label{sp}
\eer
At finite $N$ the deviation of the zeros from this critical curve is
of order $1/\sqrt N$.

\begin{figure}[t]
\centerline{\psfig{file=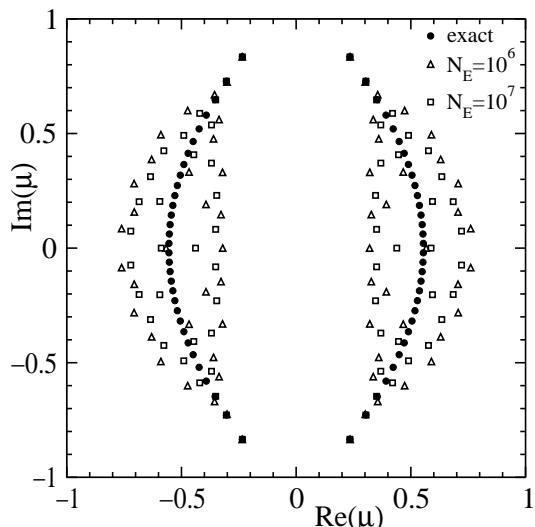,width=7cm,angle=-90}}
\caption{
Zeros from averaging over $10^6$ and $10^7$ configurations 
and exact zeros for a quenched ensemble with $N=32$.
}
\label{p32}
\end{figure}

\section{Glasgow averaging in the RMM}

The unquenched partition function for given $\mu$ and $m$ 
can be thought of as the quenched expectation value of the fermion determinant.
If the zeros of the partition function $\xi_k$ are known, we can
express the partition function as a product over its zeros
\ber
\langle \prod\limits_{k} \lp \lambda_k + \mu \rp \rangle ~=~
\prod\limits_k \lp \xi_k + \mu \rp~~.
\eer
In this identity, the zeros of the partition function appear as a kind of  
``averaged eigenvalues'' of the propagator matrix.
The Glasgow method from lattice QCD attempts to perform this averaging.
For a given matrix from the quenched ensemble, one writes out the coefficients
of the polynomial on the left-hand side.
The average coefficients are the coefficients of the right-hand side.
The main focus of this paper is to study the convergence  properties 
of the zeros to the exact ones for finite ensembles.  

The eigenvalues are in general complex. 
Because of the structure of $F(m)$, the eigenvalues occur in pairs 
$\lb \lk,-\lk^* \rb$.
Also the matrix $-C$ occurs with the same probability as $C$ in (\ref{form5}).
Therefore the ensemble average will also 
contain the pair $\lb - \lambda_k, \lambda_k^* \rb$.
Upon ensemble averaging, one can then easily show that the odd 
coefficients vanish and the remaining coefficients are real.

These simplifications should not mislead us into believing that we have safely
avoided the trouble of averaging over the complex phase of the determinant.
As we will see shortly, the problem will show up in the form of a very high 
precision needed for the coefficients in order to calculate the roots reliably.
Since the suppression achieved by averaging over the phase of the determinant
is on the order of the magnitude of the unquenched partition function,
which for $m=0$ is $\exp \lp- N (1 + \mu^2) \rp$,
it becomes exponentially difficult to achieve such precision.

The quenched ensemble is not the only way to sample the set of all matrices.
One may multiply and divide by any convenient function in the above formula.
One factor modifies the ensemble, i.e., the way the individual matrices are
generated, and the other one is used to compensate each configuration for the
modified weight.
One obvious choice is to use the unquenched ensemble at $\mu=0$.

In the next section we report the results of numerical experiments performed
using Glasgow averaging for the partition function (\ref{form5}), 
both with the quenched ensemble and the unquenched
ensemble at $\mu=0$.

\begin{figure}[t]
  \centering
  \subfigure[quenched ensemble]{
    \psfig{file=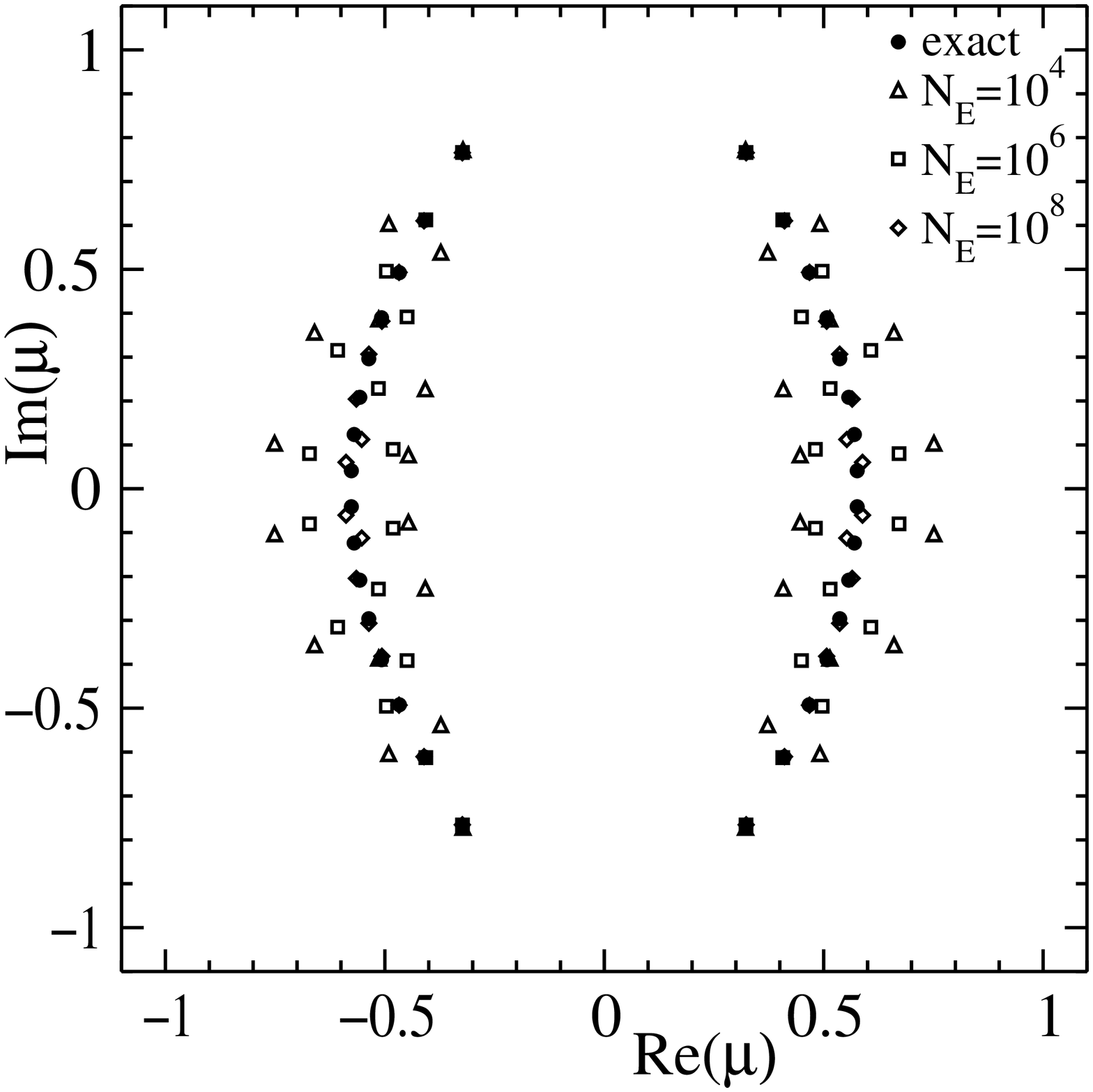,width=8cm,angle=-90}
    \label{p016}}
  \subfigure[unquenched ensemble ($N_f=1$)]{
    \psfig{file=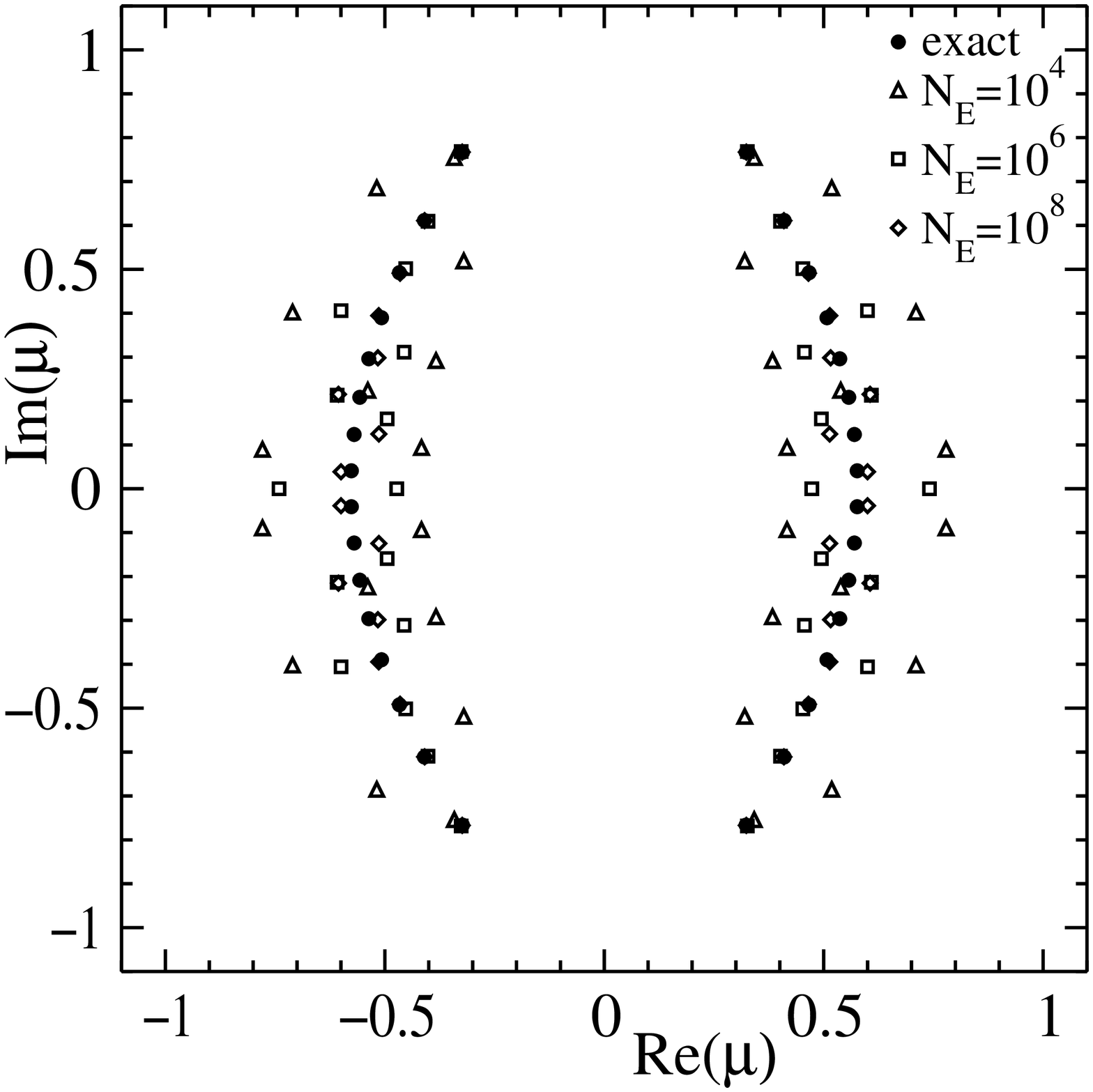,width=8cm,angle=-90}
    \label{p116}}
  \caption{Zeros from averaging over $10^4$, $10^6$ and $10^8$ configurations
           and exact zeros for ensembles with $N=16$.}
  \label{p16}
\end{figure}

\subsection{Numerical simulations}

We performed two different kinds of simulations. For $m \ne 0$, we generated
matrix elements with the Gaussian distribution given by (\ref{form5}).
We constructed the propagator matrix and obtained its eigenvalues.
For each set, the coefficients of the corresponding polynomial are calculated
and added to the average. 
For $m=0$, a much more economical procedure is possible, namely, generating
sets of eigenvalues directly, using the exact Ginibre distribution 
(\ref{N_point}).
In this case we have employed a Metropolis algorithm. 
A given set is varied using small steps and the modified set is accepted
depending on the corresponding value of the weight function. 
We have found that both cases have similar convergence properties, and
in this paper we will only report on the case $m=0$. 

In Fig.~\ref{p32} we show the zeros in the complex $\mu$ plane for a
quenched ensemble of matrices of size $N=32$ averaged over $N_E = 10^6, 10^7$
configurations.
The averaged roots do converge to the exact values as expected,
but the convergence is extremely slow.
The unconverged zeros are the ones situated closer to the real axis. 
They are situated in a cloud of a well defined shape and most of them 
are located on its edge.
As the averaging proceeds the cloud shrinks and finally disappears. 
All the roots situated outside the cloud are obtained 
correctly for a given $N$, $N_E$ combination.
This is illustrated in Fig.~\ref{p16} where we perform ensemble averaging
as large as $N_E = 10^8$.
In order to obtain more converged roots,
we had to reduce the size of the matrices to $N = 16$.
In this figure we also show results for $N_E = 10^4$ and $10^6$.
The roots next to the real axis are always the last to converge.
This is unfortunate since the value of the critical chemical potential 
(in our case $\mu_c=0.527$) is determined precisely by discontinuity
across the real axis.
The same pattern is observed for various matrix sizes $N$.
The shape of the cloud of unconverged zeros is very similar. 
However, the number of configurations $N_E$ corresponding to a given degree
of convergence increases sharply with the matrix size $N$.
One is able to determine $\mu_c$ with reasonable accuracy 
only for small values of $N$.
As we can see in Fig.~\ref{p16}, even in that case a very large
number of configurations is required.
Only after averaging over $10^8$ configurations do we get a good estimate
of $\mu_c$, but the roots are still not completely converged.

\begin{figure}[th]
  \centering
  \begin{minipage}[b]{0.45\linewidth}
    \centering
    \psfig{file=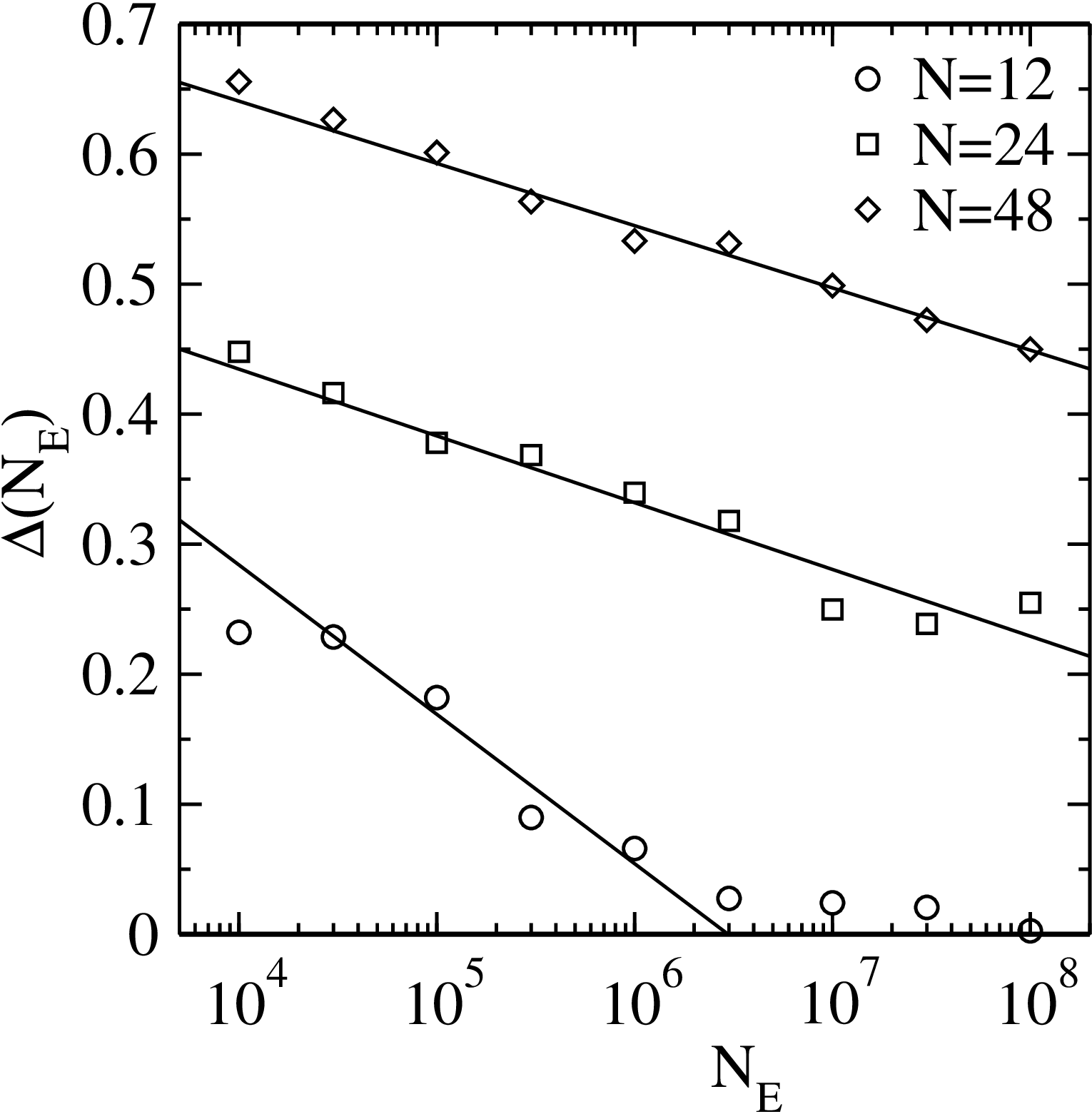,width=8cm,angle=-90}
  \end{minipage}
  \hspace{7mm}
  \begin{minipage}[b]{0.45\linewidth}
    \centering
    \psfig{file=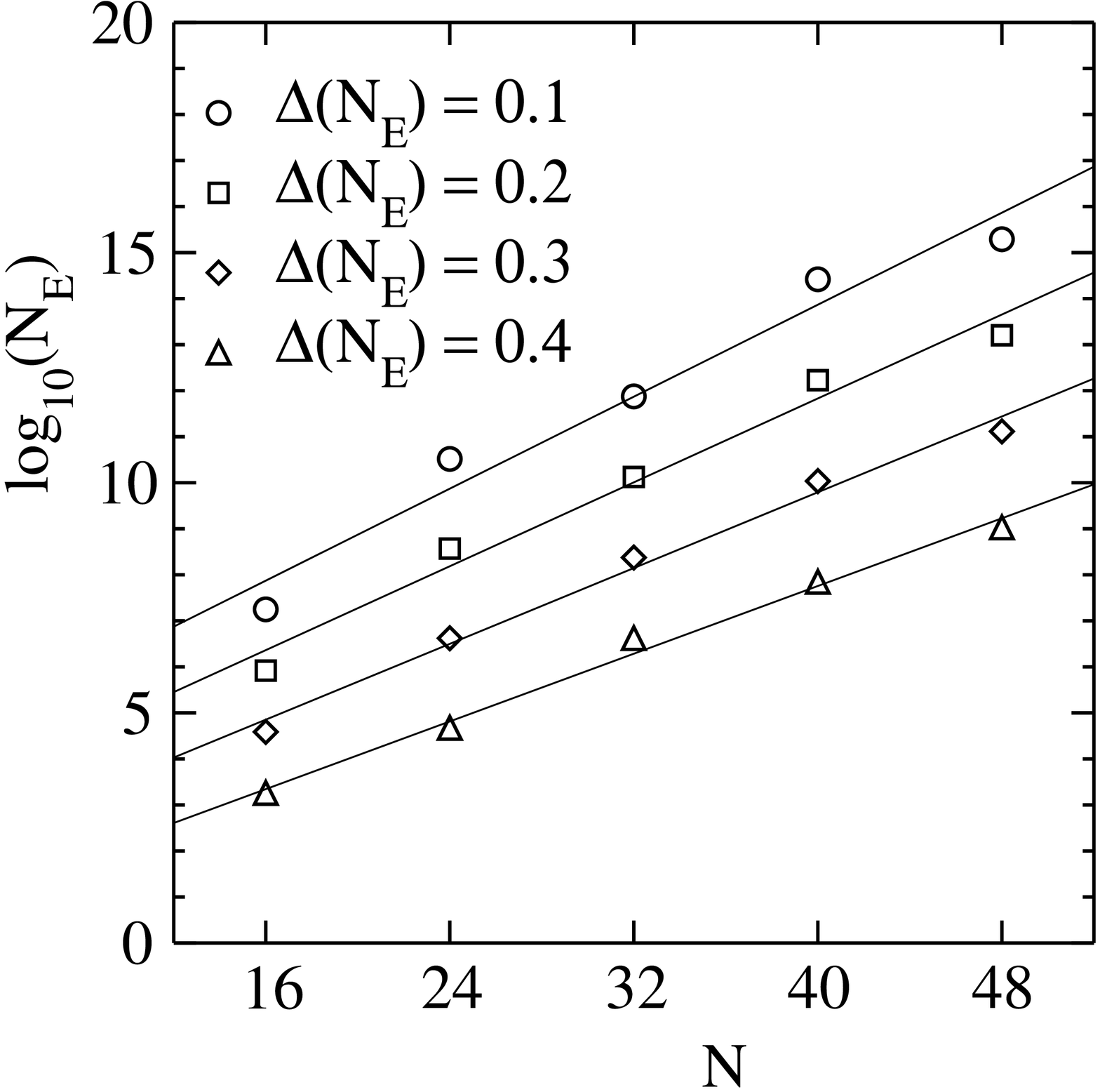,width=7.75cm,angle=-90}
  \end{minipage}\\
  \begin{minipage}[t]{0.45\linewidth}
    \centering
    \caption{The width $\Delta(N_E)$ of the unconverged cloud as a function
             of the number of configurations $N_E$. Notice that $\Delta(N_E)$
             shows a logarithmic dependence on $N_E$.}
    \label{pc}
  \end{minipage}
  \hspace{7mm}
  \begin{minipage}[t]{0.45\linewidth}
    \centering
    \caption{The logarithm of the number of configurations ($N_E$) needed to
             achieve the given error ($\Delta(N_E)$) in the zeros near the
             real axis for different sizes of matrices ($N$).}
    \label{cn}
  \end{minipage}
\end{figure}

Convergence is not improved by including the determinant at $\mu=0$ 
in the statistical weight used to generate the eigenvalues.
It was suggested previously \cite{Ha98a} that using an unquenched ensemble
at zero chemical potential might improve the efficiency of the averaging.
Our results indicate that, at best, using the unquenched ensemble has no
effect, but as it appears from Fig.~\ref{p16} the results are actually
worse than for the quenched ensemble.
An easy explanation is that a small number of configurations with 
the determinant close to zero are assigned a very large weight factor
(cf. (\ref{form1})) which has a destructive effect.
Using a ``negative'' number of flavors is another 
possibility but it is not likely to improve convergence.

We were able to run simulations which show some noticeable degree of
convergence with $N$ up to $48$.
There are many ways one could measure the degree of convergence of 
a given set of roots.
The measure we will employ is the distance $\Delta(N_E)$ between where the
two curves forming the boundary of the unconverged roots cross the real axis.
For the lower boundary we use the smallest imaginary part of the eigenvalue
near the real axis.
The upper estimate is made by running a circle through the points near the
real axis and choosing the one with the largest radius.
In both cases the determination of how close to the real axis the points should
be is made by eye, choosing a cutoff that appears to give a reasonable fit
to the boundary near the real axis.
Ideally these are the points where the contour of the cloud of unconverged
roots crosses the real axis, corresponding to a lower and an upper estimate 
for $\mu_c$ using the given ensemble. 

The dependence $\Delta(N_E)$ on the size of the ensemble, $N_E$, is illustrated
in Fig.~\ref{pc} where we show results for $N=12$, $N=24$ and $N=48$. 
The thickness of the cloud of zeros, $\Delta(N_E)$,
shows a logarithmic dependence on $N_E$.
We have included logarithmic fits for the range $N_E=10^5$ through $10^8$
($10^6$ for $N=12$).
Of course, once $\Delta(N_E)$ becomes close to zero, it stays that way. 
The slopes of the fits for $N=24,48$ vary only slightly with $N$. 

Finally, the issue of most practical interest is
how the number of configurations $N_E$ required to achieve 
a fixed value of $\Delta(N_E)$ varies with the size of the matrix. 
This was estimated by fitting lines to the available data for
$\Delta(N_E)$ versus $\log_{10}(N_E)$.
As we saw in Fig.~\ref{pc} this does give nice linear fits.
We then chose values of $\Delta(N_E)$ and calculated where the linear fits
intersected these values for each value of $N$.
Our results are plotted in Fig.~\ref{cn}.
We conclude that the number of configuration required to obtain a given
precision increases exponentially with the matrix size $N$.
This is the most important conclusion of this paper. 
In the remaining sections we will attempt to reinforce this conjecture by
studying the sensitivity of the zeros of the exact partition function to
small random perturbations of the corresponding polynomial coefficients.

\subsection{Perturbed exact zeros}

It is a well known fact that extreme care has to be taken when calculating
zeros of very high order polynomials. 
In particular, this is the case for a finite
representation of a partition function, such as the one under investigation,
where one is ultimately interested in large values of $N$. 
In our numerical experiments the polynomial coefficients are obtained as a
result of a statistical averaging process.
The averaged coefficients are approximations of the exact ones.
As the simulation proceeds, the error decreases. 

We wish to investigate the sensitivity of the zeros of the partition
function to the precision with which the coefficients are obtained.
We consider the exact polynomial and add a fixed relative error to each
coefficient:
\ber
\tilde{c}_k = c_k (1 + R_k \epsilon)~~,
\eer
where $R_k$ are random numbers between $[-1,1]$, and $\epsilon$ is a small real
positive number.  We then calculate the zeros of the polynomial 
$\sum\limits_{k} \tilde{c}_k z^k$.

\begin{figure}[t]
\centerline{\psfig{file=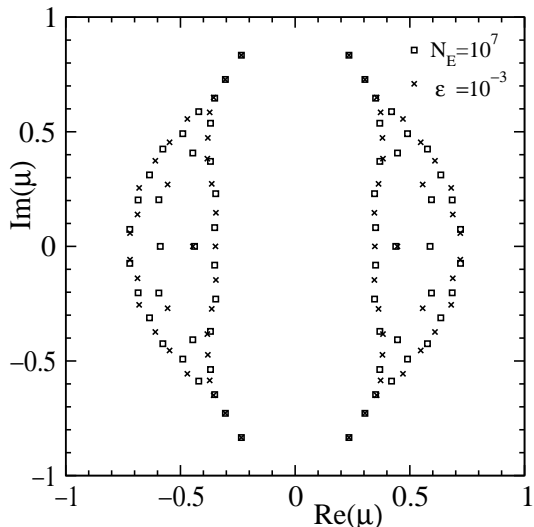,width=7cm,angle=-90}}
\caption{
Zeros from averaging over $N_E = 10^7$ configurations
and zeros of the exact polynomial perturbed with
an error factor of $\epsilon = 10^{-3}$ both for $N = 32$.
The pattern of the false zeros is practically the same for  
Glasgow averaging and the polynomial with artificial noise.
}
\label{rte32}
\end{figure}

The effect of obtaining approximate coefficients by Glasgow
averaging is quite similar to that of simply adding ``artificial noise'' 
to the exact coefficients. 
In Fig.~\ref{rte32} we plot the roots obtained for $N = 32$ via Glasgow
from $N_E = 10^7$ configurations and the roots of the exact polynomial
perturbed by noise of magnitude $\epsilon = 10^{-3}$. 
The patterns of the two sets of roots are hardly distinguishable. 

In this subsection we study the effect of such small random perturbations.
In particular, we are interested in the dependence of the precision 
$\Delta(N_E)$ (in obtaining $\mu_c$) on $N$ and $\epsilon$.
This approach has the advantage that we can consider larger values of $N$ than
in the case of direct Glasgow averaging. 
We expect that the relation between the error parameter $\epsilon$ and the number of
Glasgow configurations necessary to achieve the same accuracy is given by
the central limit theorem, $\epsilon \sim 1/ \sqrt{N_E}$.

One striking feature for larger $N$ is the extreme sensitivity of the 
roots to small perturbations.
For $N=96$, a noise factor of $10^{-18}$ already leads to 
a 20\% error in the critical value of the chemical potential.
Therefore, computing the zeros for $N=96$ is already beyond
the capability of standard double precision.\footnote{
In all our calculations we use multiprecision arithmetic,
implemented either using the GNU multiprecision package
or the one made publicly available by NASA \cite{MP}.} 

\begin{figure}[t]
  \begin{minipage}[t]{0.46\linewidth}
    \centering
    \psfig{file=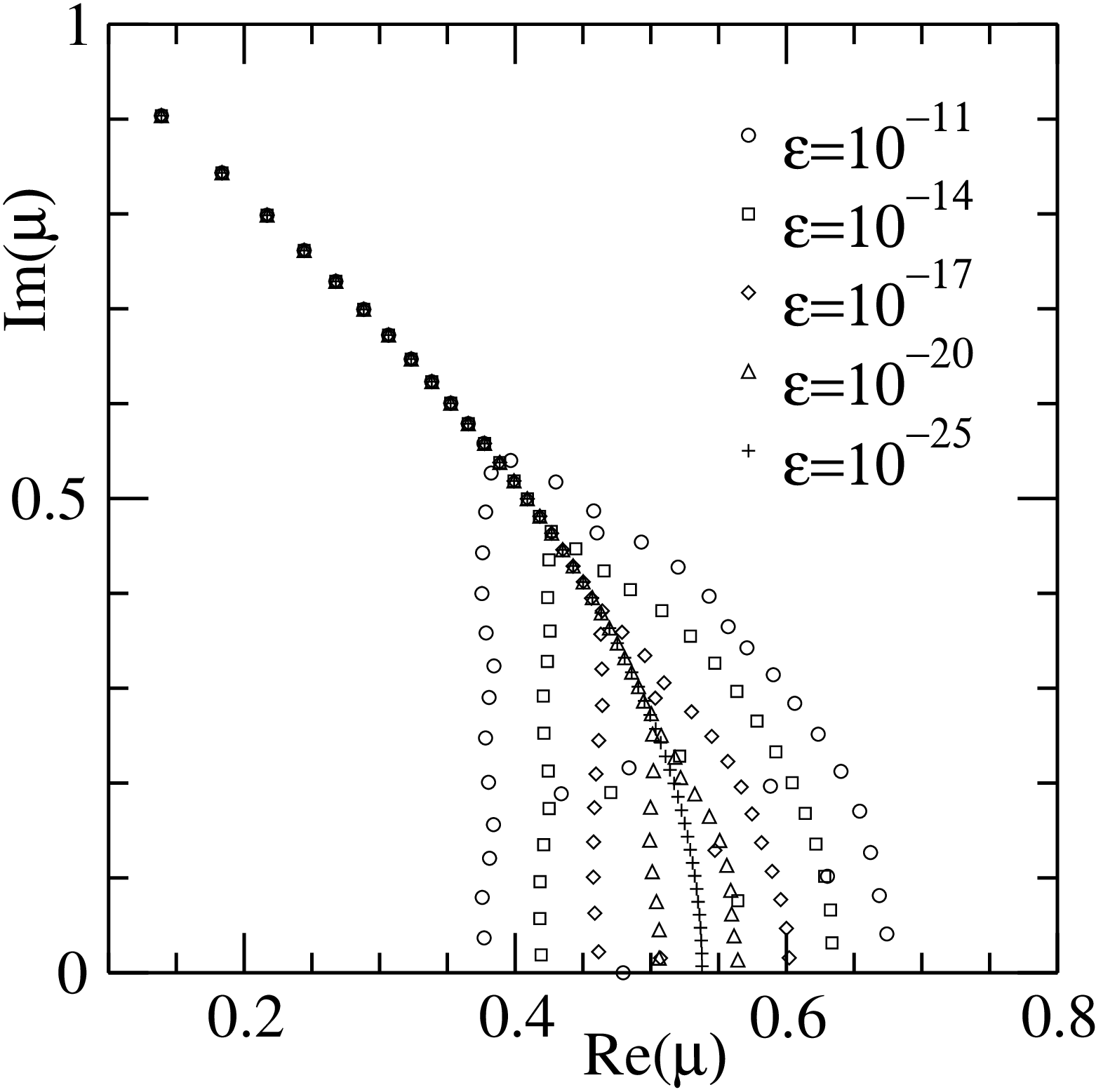,width=8cm,angle=-90}
    \vspace{4mm}
    \caption{Zeros of the perturbed RMM partition function for $N=96$ with
             several values of the error coefficient $\epsilon$.}
    \label{noisy1}
  \end{minipage}
  \hspace{3mm}
  \begin{minipage}[t]{0.46\linewidth}
    \centering
    \psfig{file=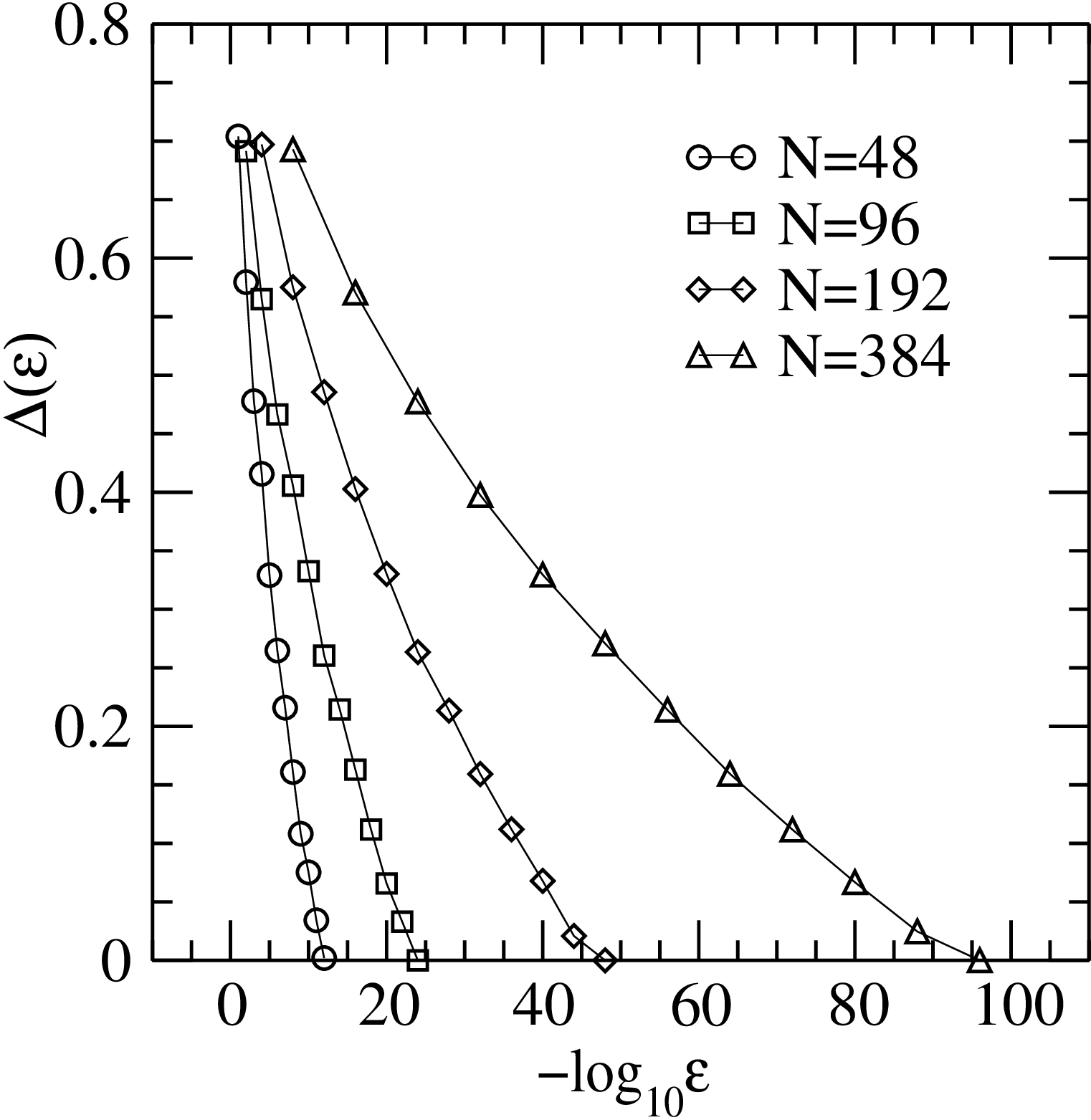,width=8.3cm,angle=-90}
    \vspace{1.4mm}
    \caption{$\Delta(\epsilon)$ for $N=48,96,192,384$.}
    \label{noisy2}
  \end{minipage}
\end{figure}

In Fig.~\ref{noisy1} we plot zeros with different degrees of artificial noise 
(see label of figure) for $N=96$.
The zeros for $\epsilon=10^{-25}$ coincide with the exact roots.
The noise factor for the remaining four sets of zeros increases with the
same factor of $10^3$ from one set to the next.
The intercepts of the edge of the eigenvalue cloud with the real axis are
practically equally spaced for the different sets of zeros. 
The precision $\Delta(\epsilon)$ appears to be a linear function of
$\log(\epsilon)$ until $\Delta(\epsilon)$ vanishes.

In Fig.~\ref{noisy2} we plot $\Delta(\epsilon)$ versus $-\log_{10}(\epsilon)$
for $N=48,96,192,384$.
For error values not too for away from convergence, we observe 
a linear relationship between $\Delta(\epsilon)$ and $-\log(\epsilon)$. 
In all cases convergence is approximately achieved when
$-\log_{10}(\epsilon) \approx N/4$.
This shows that the accuracy required to achieve a given precision
increases exponentially with $N$.
These results are consistent with results obtained with Glasgow averaging
where the relative variance of the coefficients is roughly constant and
varies as $\sim 1/\sqrt N_E$ with only a weak dependence on $N$.  

We conclude that adding random noise to the exactly known coefficients has the
same effect on the roots of the partition function as Glasgow averaging.
This confirms once more that an exponentially large number of configurations
is required to obtain a given precision. 
Extrapolating the $N$-dependence of the number of configurations needed
given by the linear interpolation in Fig.~\ref{cn} to lattice simulations
leads to extremely large numbers for even a small lattice size.
Consider for example a matrix size of $N=128$ (so $D$ in (\ref{form5})
is $256 \times 256$), 
corresponding to a $4^4$ lattice with one degree of freedom per site.
To achieve a precision of $\Delta = 0.3$ we would need approximately
$N_E = 10^{28}$ configurations, and for $\Delta=0.2$ we would need
$N_E = 10^{32}$ configurations.
A much more reasonable number of configurations can only be achieved if we
consider a small lattice of $2^4$ which corresponds to $N=24$
for Kogut-Susskind (staggered) fermions.
Here we would need $N_E = 10^6$ configurations for $\Delta=0.3$ or
$N_E = 10^8$ configurations for $\Delta=0.2$.
This exponential dependence was noted previously \cite{Ba99,Ha98a}.
In \cite{Ba99}, a signal was achieved for a lattice of $2^4$ with $10^5$
configurations but for a $4^4$ lattice only a very weak signal was found 
after $2 \times 10^6$ configurations.

\subsection{Further analysis of the convergence of Glasgow averaging}

In the previous two subsections we hope to have convinced the reader
that the phenomena accompanying Glasgow averaging are nothing but the effect
of knowing the polynomial coefficients of the partition function only with
limited precision. This follows from the fact that the pattern of the false
(unconverged) zeros is reproduced by adding small random numbers to the 
exact polynomial coefficients.

In our model the zeros close to the real axis are the last to converge.
These are also the most interesting in practice, since they
determine the critical value $\mu_c$.
It is conceivable that for certain situations in QCD (such as with finite
temperature) the roots close to the real axis are among the faster converging
ones, which would provide a glimmer of hope for the Glasgow method.

In this subsection we consider the effect of artificial noise in detail. 
We wish to understand this phenomenon as well as why the false zeros are
always concentrated in a cloud of a well defined shape.
We will find that there is a clear correlation between the magnitude of the
partition function and the stability of the zeros.

\begin{figure}[t]
\centerline{
\hspace{-0.2cm}
\epsfig{file=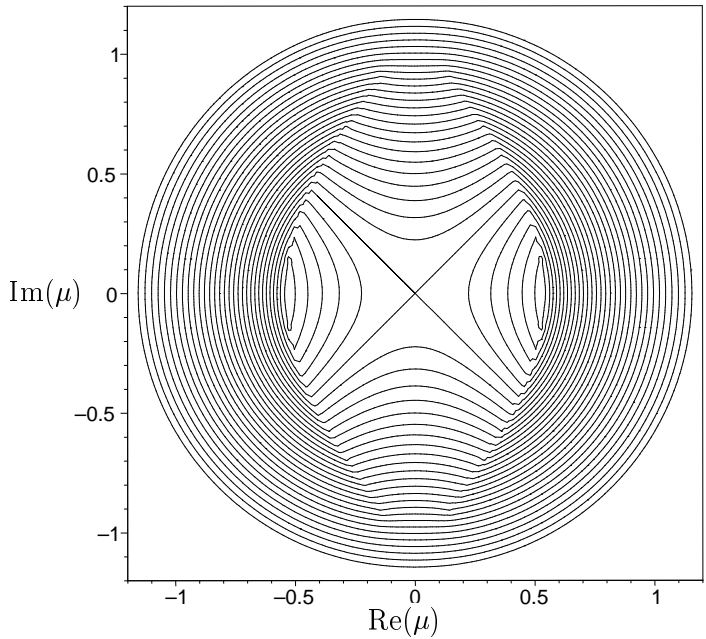,bbllx=165,bblly=312,bburx=460,bbury=630,width=8cm}
}
\caption{Level curves of the absolute values of the free energy per site
for the RMM as a function of the complex parameter $\mu$.}
\label{rmmcont}
\end{figure}

In Fig.~\ref{rmmcont} we give a topographic map of the absolute value of
the RMM partition function on a logarithmic scale as a function of $\mu$.
That is, we plot the level curves of the free energy per site given
(up to a constant) by $\ln(|Z_N|^2)/N$.
This is the natural quantity to study, since the partition function scales 
exponentially with $N$ \cite{Ha97,Ha98}. We observe a discontinuity in the
derivative of the absolute value of the partition function in the complex
$\mu$ plane. This is the locus of the zeros of the partition 
function. The deepest points are those around the (real) critical value
of the chemical potential. A quick comparison with the preceding scatter plots
should convince the reader that the higher the value of $\ln (|Z|)$ 
the more robust are the zeros in that neighborhood. 
However, the level curves do not match the contour of the cloud of 
false zeros. The situation a is little more subtle as we will see below.

\subsubsection{The noisy partition function}

It is useful to separate the approximate (``noisy'') partition function into
the exact one and the part due to the error in the coefficients,
\ber
Z_{\rm tot}(\mu)~=~Z_{0}(\mu) + Z_{\rm err}(\mu)~~.
\eer
The partition function $Z_{\rm err}(\mu)$ is a polynomial whose coefficients
are the differences between the exact coefficients and the approximate ones,
\ber
Z_{\rm err}(\mu)~=~
\sum\limits_{k=0}^N \delta c_k \mu^k~~;~~\tilde{c}_k = c_k + \delta c_k~~.
\eer
The differences are generated as $\delta c_k = \epsilon R_k c_k$ 
in the `artificial noise' case.
Our initial assumption was that for true Glasgow averaging the relative error
in the coefficients is approximately the same for all zeros.
This assumption was reinforced by the similarity between the 
two types of results discussed in previous subsections. In the following,
we will discuss mainly the `artificial noise' partition function.

In terms of $Z_{\rm err}(\mu)$,
an explanation of the qualitative picture we have observed is the following. 
In the regions
of the complex $\mu$ plane where $|Z_{\rm err}(\mu)| \ll |Z_0(\mu)|$ 
one may ignore the $Z_{\rm err}(\mu)$.
The roots of the total partition function located in these regions
coincide to a good degree with the corresponding exact roots. 
On the contrary, in regions 
where $Z_{\rm err}(\mu)$ dominates, the roots are determined by the latter and
their general pattern has no similarity to their exact counterparts. 
This is consistent with the fact that the robust zeros are the ones 
situated in the region with higher $\ln(|Z_0|)$. 
The shrinking of the cloud of false zeros with decreasing $\epsilon$ is a
consequence of the corresponding decrease in the magnitude of $Z_{\rm err}$. 

Throughout most of the complex $\mu$ plane, $Z_0$ is clearly larger than
$Z_{\rm err}$.
However, near the roots of $Z_0$ the error part has its chance to dominate.
This is because the value of the exact partition function is the
result of a major cancellation in the region where $Z_0$ is 
well approximated by $\exp(-\mu^2 N)$.
The sum of the (finite) series is much smaller 
than the general term, which is typically
of order $1$. By multiplying each term of this series with 
a random number we spoil the 
cancellations. The sum of the perturbed series may therefore 
be significantly larger in absolute
value than the exact sum.

\begin{figure}[t]
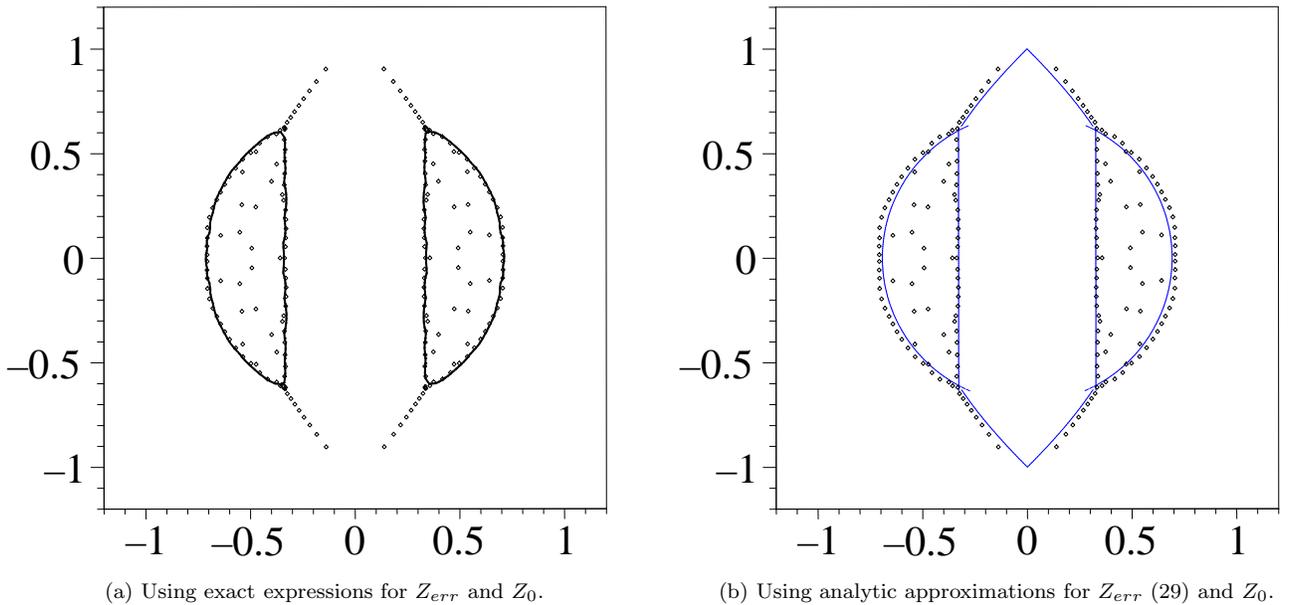

  \centering
  \subfigure[Using exact expressions for $Z_{err}$ and $Z_0$.]{
    \begin{minipage}[t]{0.45\linewidth}
      \hspace{4mm}
      \psfig{file=last1.ps,width=8cm,angle=-90}
      \vspace{6mm}
    \end{minipage}
    \label{errcont1}}
  \hspace{4mm}
  \subfigure[Using analytic approximations for $Z_{err}$
             (\ref{erran}) and $Z_0$.]{
    \begin{minipage}[t]{0.45\linewidth}
      \hspace{4mm}
      \psfig{file=last2.ps,width=8cm,angle=-90}
      \vspace{6mm}
    \end{minipage}
    \label{errcont2}}
  \caption{Zeros for $N=96$ and artificial error $\epsilon=10^{-8}$, with the
           curve $|Z_{err}/Z_{0}|=1$.}
  \label{errcont}
\end{figure}

\subsubsection{Sensitivity of the zeros}

One arrives at similar conclusions by studying the 
infinitesimal variation of the roots.
Let $\mu_k$ be the exact roots (so that $Z_0(\mu_k)=0$) and 
let $\mu_k + \delta \mu_k$ be the roots of the total
noisy partition function. 
From our decomposition, we have 
\ber
Z_0(\mu_k + \delta \mu_k) 
+ Z_{\rm err}(\mu_k + \delta \mu_k)~=~0~~~\rightarrow~~~
\frac{|Z_{\rm err}(\mu_k + \delta \mu_k)|}{|Z_0(\mu_k + \delta \mu_k)|}
=1~~.
\label{cont}
\eer
Since the modified roots are not zeros of $Z_0$ or $Z_{err}$, the above
equality is not fulfilled trivially.
It indicates that the false zeros should be located in the transition region
where $|Z_0|$ and $|Z_{err}|$ are comparable, i.e., on the border of the
region where $Z_{err}$ dominates, rather than scattered inside it.
In Fig.~\ref{errcont} we show how the contour of the cloud of false zeros is 
obtained using the formula above.

We can also make a quantitative estimate of how well converged a given
root is.
If we expand the first half of (\ref{cont}) we get
\ber
Z_0(\mu_k) + \delta\mu_k Z_0^\prime(\mu_k) +
Z_{\rm err}(\mu_k) + \delta\mu_k Z_{\rm err}^\prime(\mu_k) \approx 0.
\eer
The first term is simply zero.
The last term we will neglect as being of higher order in $\delta \mu_k$.
The remaining two terms can be rearranged to give
\ber
\delta \mu_k ~\approx~ - \frac{Z_{\rm err}(\mu_k)}{Z_0^\prime(\mu_k)}.
\eer
In other words, the variation of a given root $\mu_k$ 
is proportional to the ratio of $Z_{err}$ to $Z_0^\prime$.
If this quantity is negligible, the root is close to the exact one.
If the ratio is close to $1$ or larger, the shift in $\mu_k$ is large,
and the root is not obtained correctly.
Therefore we can say that in the region where
$|Z_{\rm err}| \ll |Z_0^\prime|$, the roots
of $Z_{tot}$ are reliable.

\begin{figure}
  \begin{minipage}[t]{0.46\linewidth}
    \centering
    \epsfig{file=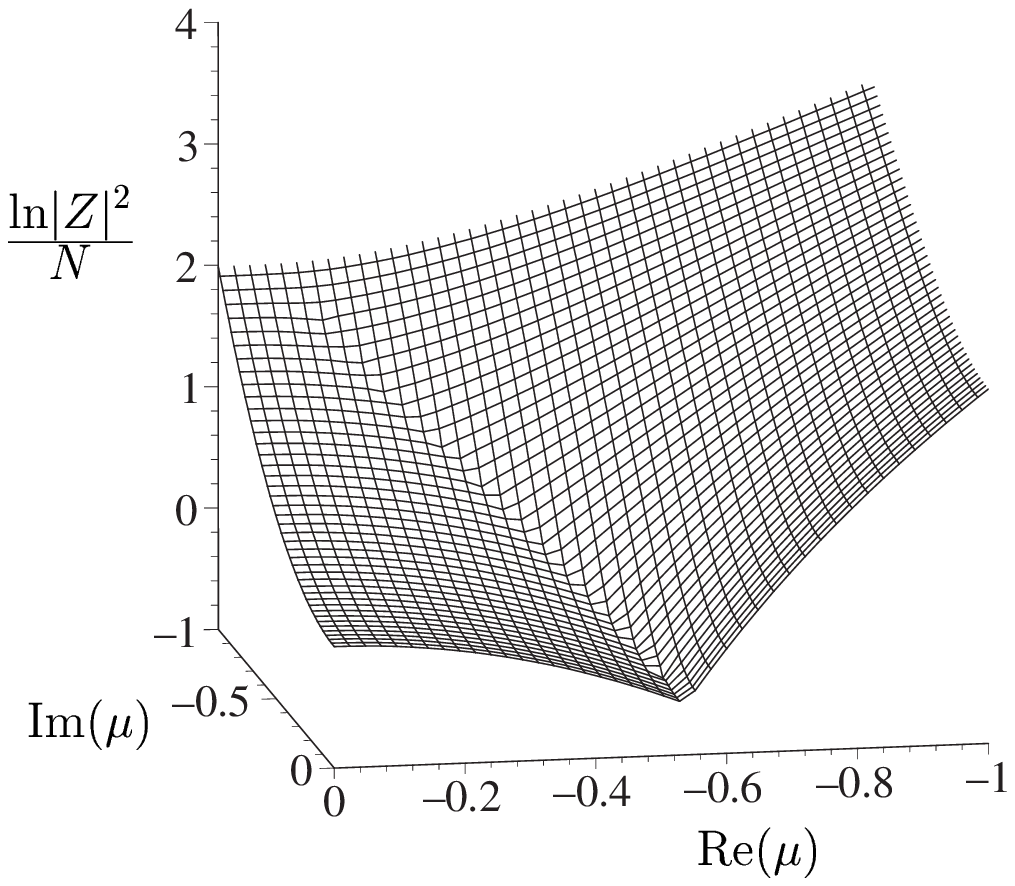,bbllx=140,bblly=350,bburx=450,bbury=610,width=8cm}
    \caption{The absolute value of the free energy per site for the RMM with
             $N=96$.}
    \label{rmmfe}
  \end{minipage}
  \hspace{5mm}
  \begin{minipage}[t]{0.46\linewidth}
    \centering
    \epsfig{file=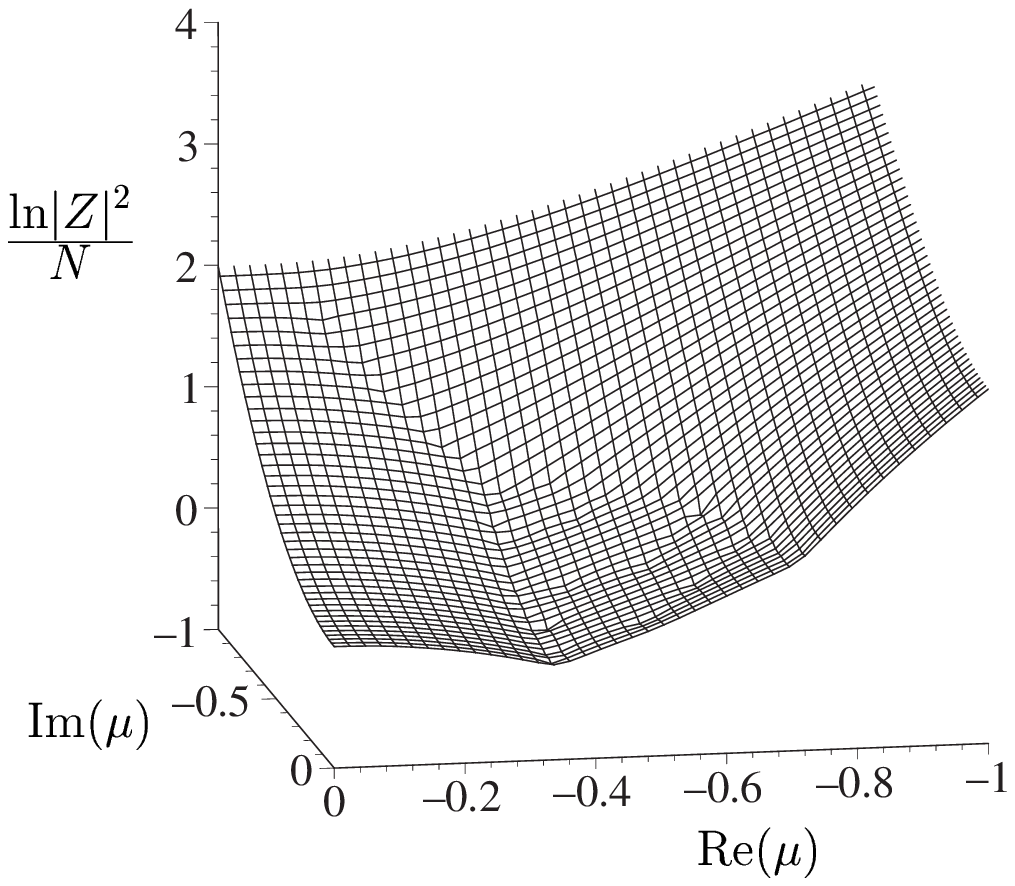,bbllx=140,bblly=350,bburx=450,bbury=610,width=8cm}
    \caption{The absolute value of the free energy per site for $N=96$ with
             noise $\epsilon = 10^{-8}$}
    \label{noisefe}
  \end{minipage}
\end{figure}

\subsubsection{Error piece as additional phase}

We may check our arguments in the previous sections by comparing the magnitudes
of the exact, the total and the noise partition functions (or free energies)
in the complex $\mu$ plane.
Fig.~\ref{rmmfe} shows the absolute value of the free energy per site,
$\ln(Z_0)/N$ in one quadrant of the complex $\mu$ plane. 
The location of the exact zeros is on the cusp which runs from $\mu=-i$ to
$\mu=0.527$. 

\begin{figure}[t]
\centerline{
\hspace{-2.0cm}
\epsfig{file=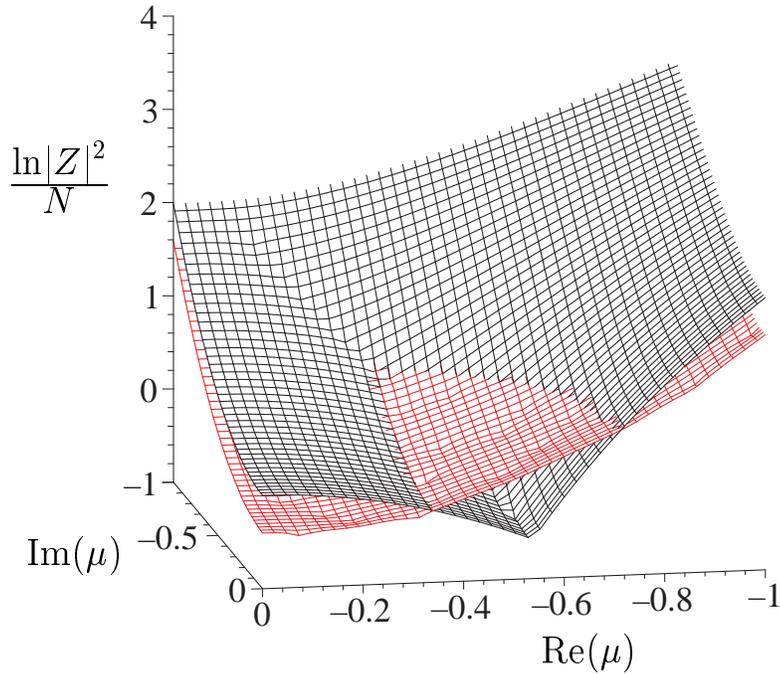,bbllx=140,bblly=350,bburx=450,bbury=610,width=11cm}
}
\caption{The exact (the same as in Fig.\ref{rmmfe})
and the error partition functions for $N = 96$ and
$\epsilon = 10^{-8}$ }
\label{exerrfe}
\end{figure}

In Fig.~\ref{noisefe} we have the same type of plot now for the total partition
function with a noise factor $\epsilon = 10^{-8}$.
The cusp in the noisy result bifurcates. 
The exact and the noisy surfaces coincide exactly except for the region
between the two new branches of the cusp, where the noisy partition function
is larger. 
The new cusps coincide with the locus of the false zeros. 
A few false zeros are also scattered inside the region.
It is clear from Fig.~\ref{exerrfe} where we plot the surfaces corresponding to
$Z_0$ and $Z_{\rm err}$ simultaneously that the new cusps are located at the
intersection of the free energy surfaces.
The exact piece dominates everywhere except for inside the bifurcation,
where the error piece dominates.  
The $\mu$ dependence is so steep for both of them that the smaller piece
becomes negligible very fast as one moves away from the intersection line.

When we discussed earlier the properties of polynomials as partition functions,
we associated the different analytic functions which are approximated by the
polynomial in different regions of the complex parameter space with the phases
of the partition function. 
We mentioned that the zeros are typically located along the phase transition
lines, since within any given phase the partition function
is a smooth analytic function which does not vanish.
The partition function $Z_{\rm err}$ has a quasi-analytic
behavior similar to that of the partition function itself.
The bifurcation of the line of zeros can be interpreted as the presence of
an additional spurious `phase' in the partition function, namely, 
the region where the error piece dominates.
Then the fact that the false zeros are located on the 
$|Z_{\rm err}|~=~|Z_0|$ line is natural.

\subsubsection{Analytic approximation of the error partition function}

\begin{figure}[t]
\centerline{
\hspace{-2.0cm}
\epsfig{file=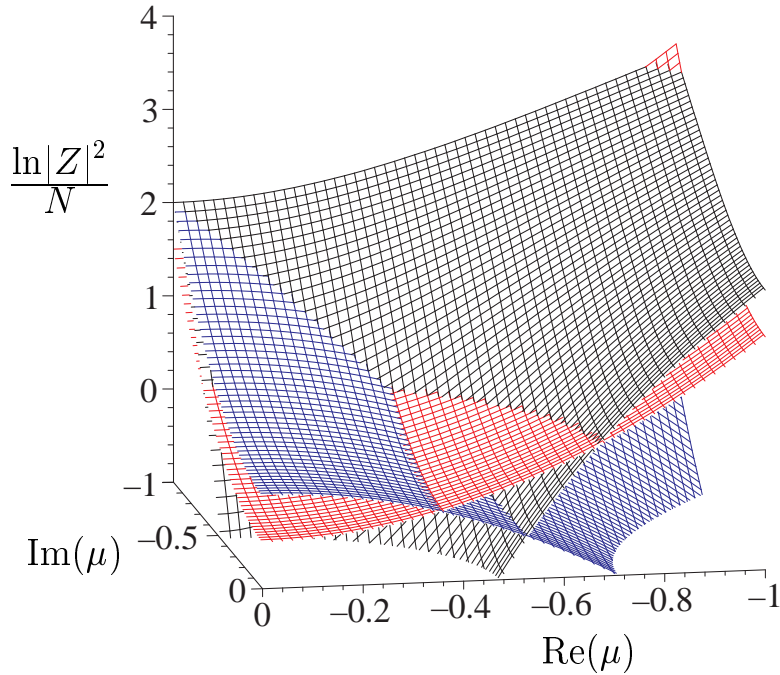,bbllx=140,bblly=350,bburx=450,bbury=610,width=11cm}
}
\caption{The exact and the error partition functions computed analytically.}
\label{exerran}
\end{figure}

The question is whether the error partition function can be 
at all approximated by an analytic function.
In the present case the answer is very simple.
Let us consider the error partition function,
\ber
Z_N^{\rm error}(\mu) ~=~ 
\epsilon \sum\limits_{k=0}^N R_k \frac{ (- N \mu^2)^k}{ k! }~~.
\eer
It is the same truncated inverse exponential we have discussed, only now each
term is multiplied by a random number $R_k$ of order $1$.
The terms in the original series conspire to achieve a major cancellation,
of order $e^{- N {\rm Re} \mu^2 }$. 
The random numbers spoil this, and each term in the series is left to fend
for itself, and the sum is dominated by the largest terms.
From the ratio of two consecutive terms in the sum,
\ber
-\frac {N\mu^2}{k}
\eer
we conclude that the largest term (in absolute value) is 
the one with $k = k_{\rm max} = | \mu^2 N |$.
If $|\mu^2| > 1$ then the largest term is the one with $k=N$.
Of course the surrounding terms must have a significant contribution, 
but the end result should still be proportional to this largest term. 
For $|\mu| < 1$ our estimate is therefore
\ber
| Z_N^{\rm noise}(\mu) |~ \approx ~ {\cal C} \frac{k_{\rm max}^{k_{\rm max}}}
{k_{\rm max}!} \sim \frac{e^{N|\mu|^2}}{|\mu| \sqrt N}
\label{erran}
\eer
In Fig.~\ref{exerran} we plot the absolute value of the two surfaces
corresponding to the continuum limit, $e^{- \mu^2 N}$ and $e^{N} \mu^{2N}$,
and the one corresponding to our estimate of the error partition function
given above.
The intersections of the three surfaces follow the 
pattern of the corresponding zeros.
We also checked that (\ref{erran})
approximates the error partition function well.

As an added bonus, we can now explain the scaling with $N$ of the 
required precision.
We found that the error partition function, i.e., the exact partition function
whose coefficients have been multiplied by random numbers of order $1$,
is well approximated by an analytic expression. 
This expression shares an important property of the true
partition function, namely, {\it it scales exponentially with $N$}.
\ber
\frac{1}{N_1} \ln( Z_{N_1}^{exact}(\mu)) ~ 
\approx ~  \frac{1}{N_2} \ln(Z_{N_2}^{exact} (\mu))~~;~~
\frac{1}{N_1} \ln(Z_{N_1}^{err}(\mu) )~ \approx ~  
\frac{1}{N_2} \ln(Z_{N_2}^{err} (\mu))
\eer
The magnitude of the error partition function is also controlled by the factor
$\epsilon$ which mimics the precision to which the coefficients are
approximated in the averaging process.
The locus of the false zeros is controlled by the relative 
magnitude of the true partition function and the error partition function.
To have the same pattern of zeros, we must also scale $\epsilon$ exponentially
with $N$, $\epsilon(N) = \alpha^{-N}$ .

\section{Conclusions}

We have investigated Glasgow averaging using a random matrix model at
nonzero chemical potential.
We have found that in a quenched ensemble the method converges, but that it
requires an exponentially large number of configurations. 

The roots of the averaged polynomial are initially distributed similarly
to the eigenvalues of individual configurations.
As the averaging proceeds, the roots approach their exact values.
After averaging over a finite number of configurations,
the roots clearly separate into two groups.
Some roots are close to the corresponding exact ones.
Typically, these are the zeros far from the real axis.
The remaining roots are situated in a cloud
around the intersection of the real axis and the locus of the exact zeros.
The zeros outside the cloud are practically exact, while those inside and
on the boundary of the cloud are very badly determined.
They cannot be traced to individual exact zeros. 
The shape of the cloud is similar for different matrix sizes.
It shrinks as more configurations are taken into account. However it
only shrinks as a logarithmic function of the total number of
configurations. 

By interpolating the number of configurations needed to reach a given
precision for several matrix sizes, we were able to estimate the dependence
of the number of configurations needed for a given matrix size.
Our conclusion is that the number of configurations needed to reach a given
level of precision grows exponentially with the size of the matrix.

The results of the corresponding unquenched simulations, using an ensemble
generated with $N_f=1$ and $\mu=0$ are similar to the quenched ones. 
The unquenched ensemble may have been helpful in improving the overlap 
between the simulation and the `true' ensemble corresponding to fixed
nonzero $\mu$.
However, this would still necessitate simulating an ensemble at nonzero $\mu$,
which is precisely what we were trying to avoid.
The exponential statistics observed by us are more likely to be the
signature of the sign problem itself, i.e., the magnitude of the
cancellation brought about by the varying phase of the determinant.
Hence the Glasgow method is unable to surmount the sign problem.
However, in a problem where the latter is absent, such as $SU(2)$ simulations
\cite{Ha99},
the Glasgow method -- even quenched -- should be helpful.
It would be interesting to see how the overlap issue manifests itself
in this case.

We obtained results very similar to those of Glasgow averaging by perturbing
the exact coefficients in the RMM partition function polynomial.
We studied empirically the dependence of the reliability of the zeros on the
precision with which the coefficients are known and on the size $N$ of the
model matrix for larger matrices.
Just like in the case of averaging, this dependence is exponential.
That is, given a desired error bound for the zeros, the necessary
precision on the coefficients grows exponentially with $N$.
The extrapolation of this dependence to Glasgow averaging translates into
exponentially large statistics, since the precision on the coefficients
should be proportional to the square root of the number of configurations.

There is a correlation between the phenomenon of slower or faster
converging zeros and the magnitude of the continuum partition function.
The zeros that converge slowly are in the region where the 
partition function is suppressed.
Large cancellations require better precision, hence more statistics.
More formally, the effect of perturbing the coefficients can be understood
as the addition of an extra (error) polynomial to the true partition function.
This extra piece is found to scale exponentially with $N$, just like the true
partition function.
The effect on the phase structure can be seen as the introduction of a spurious
phase, which replaces the true ones in the regions of parameter space where 
the error partition function dominates.
The zeros in these regions follow the modified phase boundaries. 
The scaling property of the error partition function explains the need for
exponential statistics in Glasgow averaging.

\bigskip

The negative result regarding unquenched simulations 
at $\mu=0$ is perhaps disappointing. 
It indicates that the quenched ensemble 
and the unquenched ensemble at $\mu=0$ are equally
relevant to the behavior of the model close to the critical $\mu$.
The upside of the failure of the unquenched ensemble 
at $\mu=0$ is that 
in future applications of the Glasgow method
it might be worth trying to use a quenched
ensemble. We remind the reader that using the quenched
ensemble is not equivalent to the quenched approximation.

Another glimmer of hope is in the phenomenon of fast converging zeros.
In the present random matrix model the zeros that determine the critical
$\mu$ are the last to converge.
From our analysis there is no indication that the sensitive zeros 
are generally those close to the real (physical) axis.
There is a possibility that in QCD or in other interesting non-Hermitian
models the critical parameter values are determined by the robust zeros. 
It is hard to make any statement in this respect from the currently available
QCD data \cite{Ba97}.
Perhaps a schematic model with features closer to QCD in terms of eigenvalue
distribution in the complex $\mu$ plane would help clarify this issue.
In general, it would be interesting to see how much of the analysis in
this work regarding zeros of approximately known polynomial partition
functions applies to other models.

\section{Acknowledgements}

We thank I.M.~Barbour, S.~Hands, E.~Klepfish, M.P.~Lombardo 
and S.E.~Morrison for fruitful discussions.
R.D.~Amado and M.~Pl\"{u}macher are thanked for several critical readings 
of the manuscript.

This work was supported in part by 
NSF grants PHY-98-00098 and PHY-97-22101 as well as
the DOE grant DE-FG-88ER40388. 
Part of the calculations presented in this paper have been carried out 
at the National Scalable Cluster Project Center at the University of 
Pennsylvania, which is supported by a grant from the NSF. 
The multiprecision calculations were performed using the GNU multiprecision
package or the package made available by NASA \cite{MP}.

\end{document}